\documentstyle[12pt]{article}
\newcommand{\be}{\begin{equation}}
\newcommand{\ee}{\end{equation}}
\newcommand{\ba}{\begin{eqnarray}}
\newcommand{\ea}{\end{eqnarray}}
\newcommand{\bb}{\begin{thebibliography}}
\newcommand{\eb}{\end{thebibliography}}
\newcommand{\bi}[1]{\bibitem{#1}}

\newcounter{tmpeqn}
\newcommand{\arabiceqn}{\setcounter{tmpeqn}{\value{equation}}
\stepcounter{tmpeqn}\setcounter{equation}{0}
\renewcommand{\theequation}{\mbox{\arabic{tmpeqn}.\arabic{equation}}}}
\newcommand{\alpheqn}{\setcounter{tmpeqn}{\value{equation}}
\stepcounter{tmpeqn}\setcounter{equation}{0}
\renewcommand{\theequation}{\mbox{\arabic{tmpeqn}\alph{equation}}}}
\newcommand{\reeqn}{\setcounter{equation}{\value{tmpeqn}}%
\renewcommand{\theequation}{\arabic{equation}}}

\begin{document}

\begin{center}
\vspace*{0.8cm}
{ \Large \bf OVERBARRIER RESONANCES AS SOLUTIONS OF
 SET INHOMOGENEOUS SHR\"{O}DINGER EQUATIONS }\\
\vspace*{0.7cm}
{ \large Il-Tong Cheon\footnote{Fax: 82-2-392-1592, \ E-mail:
itcheon@phya.yonsei.ac.kr \ \
 Tel: (02)361 - 2610 } } \\
\vspace*{0.3cm}
{ \it Department of Physics, Yonsei University, Seoul 120 -- 749, Korea } \\
\vspace*{0.5cm}
{ \large  G.Kim \\ }
\vspace*{0.3cm}
{ \it Institute of Nuclear Physics, Tashkent, Uzbekistan }  \\
{ \it Department of Physics, Yonsei University, Seoul 120 -- 749, Korea } \\
\vspace*{0.5cm}
{ \large   A.V. Khugaev \\ }
\vspace*{0.3cm}
{ \it Institute of Nuclear Physics, Tashkent, Uzbekistan }  \\
\vspace*{0.8cm}
\end{center}

\begin{center}
{ \Large Abstract }
\vspace*{0.3cm}
\end{center}
\begin{center}
\parbox[h]{13.cm}{
In the paper the Schr\"odinger equation for quasibound resonance state with
complex energy is considered. The system of inhomogeneous differential
equations is obtained for the real and imaginary parts of wave function.
On the base of known solution of corresponding homogeneous equation,
the inhomogeneus system is solved with help of iteration procedure.
The single-particle neutron $2p$-state in the Woods - Saxon potential is
analyzed for $^{13}C$ nucleus.}
\vspace*{0.5cm}
\end{center}

\begin{center}
{ \large\bf I. INTRODUCTION }
\vspace*{0.3cm}
\end{center}
 \ \ \ { \large Many papers, published since 1980 and devoted to
investigation of
 the structure of nuclei with $A=13$, show that for description of
 the ground states of these nuclei, it is necessary to take the $2p$ - shell
 into account in a ground state [1 - 17].
%\footnote{ Of course, it is impossible to write all works, which were
% dedicated to this topics.}
 Recentiy nuclear structure calculation appeared in [11], which described nuclei
 with $A=4-16$ within a full $(0+2)\hbar\omega$ shell model space, i.e. configuration
 mixing, is not only restricted to the $1p$ - shell, but includes contributions from
 the $2s$-, $1d$-, $2p$- and $1f$-shells as well. These wave functions provide a good overall
 description of the relevant energy spectra.
 Furthermore, from the analysis of the $M1$ form factor of electron elastic
 scattering on the nucleus $^{13}C$ [3 - 11,14], it is followed that
 the $2p$ - shell is the most important one among the $2s$ -, $1d$ -, $2p$ -
 and $1f$ - shells. The similar conclusion regarding $2p$ - shell admixtures
 was obtained for the processes with pion [12,13,15]: pion photoproduction,
 radiative pion capture on $^{13}C$ and pion single charge exchage on
$^{13}C$.

 If we choose as an average field the oscillator potential, parameters of
 which are obtained in accordance with the binding energy of nucleon in
 $1p$ - shell, the binding energy of nucleon in $2p$ - shell becomes positive.
 Thereby, at the value $b\ =\ 1.663\ fm$ for oscillator parameter and
 $E_{bind}(1p)$ = $-4.947\ MeV$ for the neutron binding energy on $1p$ -
shell
 of $^{13}C$, we find the $binding\ energy$ in $2p$ - shell
 $E_{bind}(2p)$ = $25.016\ MeV$ in the same oscillator well. (Somewhat less
 value $E_{bind}(2p_{1/2})$ = $14.65\ MeV$ was obtained in work [18] from
 the analysis of the spectra $^{12}C,\ ^{13}C$ and $^{11}C$ nuclei.) And if
 the $2p$ - state with such positive energy is $bound$ in the oscillator
 potential because of the depth of oscillator well being infinity, then for
 the realistic potential of Woods - Saxon, the $2p$ - state represents itself
 as an overbarrier resonance [19]. One the other hand, the $2s$ -,
 $1d$ - states are subbarrier Gamov resonances, which can be considered,
 for example, by $R$ - matrix theory [20], or in the framework of the
potential
 model with taking into account the continuum [21] into account. Regarding to
 the overbarrier resonances, such resonances do not exist in the complete
 theory as far as we know. At present the quasi-classical approach [19] and
 the complex scaling method [23] are developed for the calculation of
 characteristics of such resonances. Therefore, the determination of the wave
 function, the energy and the width of the overbarrier resonance in the real
 nuclei with the realistic average field is a nontrivial problem.
 This paper is devoted to find a solution of this problem.

\begin{center}
\vspace*{0.5cm}
{ \large\bf II. FORMALISM }
\vspace*{0.3cm}
\end{center}

\ \ \
 For determination of the resonance energy $E_{res}$ and the width
 $\Gamma $ of the resonant $2p$ - state, it is necessary to solve
 the time-dependent Schr\"odinger equation [24]:

\ba
 i\hbar\frac{\partial \Psi(\vec r,t)}{\partial t} = \left [
-\frac{\hbar^2}{2\mu}\Delta + U(r) \right ] \Psi(\vec r,t).
\ea

And if we use the substitution $\Psi(\vec r,t) = \tilde \varphi(\vec
r)exp{\left (-i\frac{\tilde E}{\hbar}t\right ) }$,
then this partial differential equation can be converted into an ordinary
one:

\ba
 \left [ \tilde E+\frac{\hbar^2}{2\mu}\Delta - U(r) \right ] \tilde
\varphi(\vec r) = 0.
\ea

 Taking into account spherical symmetry of the potential $U(r)$ and
 substituting the expression $ \tilde \varphi_{nlj}(r) =  \chi_{nlj}(r)/r $
 for radial part of wave function, we find:

\ba
\left [ -\frac{\hbar^2}{2\mu}\frac{d^2}{dr^2}+\frac{\hbar^2}{2\mu }
\frac{l(l+1)}{r^2}+U(r) - \tilde E_{nlj}\right ] \chi_{nlj}(r) = 0,
\ea
 where $\mu$ is the nucleon reduced mass, $l$ and $j$ are orbital and
 total angular momenta of the nucleon; n is the principal quantum number,
 which denotes a number of knots in the radial wave function and is connected
 with a spectroscopic principal quantum number $n_c$ given by $n_c = n + 1$.
 For the potential $U(r) = V(r) + V_{S.O.}+V_{Coul.}$, we have the following
 expressions [16]:\\
 the Woods - Saxon potential
\ba
 V(r)=-V_{0}\left \{ 1+\exp{ \left ( \frac{r-R_{0}}{a} \right ) } \right
\} ^{-1},
\ea
and the spin - orbital interaction
\ba
V_{S.O.} = -\kappa\cdot ( \vec \sigma \cdot \vec l ) \cdot
\frac{1}{r}\frac{d}{dr}V(r),
\ea
where
%@
\begin{equation}
\begin{displaystyle}
(\vec \sigma \vec l ) =
\left\{
\begin{array}{ccc}
l,& j=l+\frac{1}{2},& \\
& & \\
-(l+1),& j=l-\frac{1}{2}.
\end{array} \right.
\end{displaystyle}
\end{equation}
The nucleus radius is determined as $R_{0}=r_{0}A^{\frac{1}{3}}$.
Here $r_0$ is the nucleon radius, $a$ is the diffuseness of potential and
$\kappa = 0.263\ fm^2$ is the parameter for the spin - orbital interaction.
The depth of potential $V_{0}$ is calculated in accordance with
the experimental value of the single-particle binding energy $E_{bind}=4.947MeV$
for the neutron on $1p$ - shell.

Although the Coulomb interaction does not need to be involved within the present framework of
model description of $^{13}C$, it should participate to the case of the proton playing a role.
It is given in the form
\begin{equation}
\begin{displaystyle}
V_{Coul.}(r) = \frac{(Z-1)e^{2}}{r}\cdot
\left\{
\begin{array}{ccc}
\frac{3r}{2R_{0}}-\frac{1}{2}\left ( \frac{r}{R_{0}} \right )^3,& r<R_{0}& \\
& & \\
1,& r\geq R_{0}.
\end{array} \right .
\end{displaystyle}
\end{equation}
The solution of equation (3) will be searched with
\ba
\chi_{nlj}(r) = u_{nlj}(r) + iv_{nlj}(r).
\ea
As $r \to \infty$, this solution must satisfy the following boundary
condition:
\ba
\chi_{nlj}(r) \longrightarrow \left (-\frac{i}{2}\right )\exp{\{i(\tilde
kr - \eta \ln{2}\tilde kr - \frac{l\pi }{2}+ \delta_{l})\}},
\ea
with $\delta_{l}=arg\Gamma (l+1 +i\eta )$, where $\eta$ is Coulomb
parameter.\\

From eq (9), we have
\begin{equation}
\begin{displaystyle}
\lim_{r \to \infty}
\frac{d\ln\chi_{nlj}(r)}{dr} = i\tilde k = \frac{i\sqrt{2\mu\tilde
E_{nlj}}}{\hbar},
\end{displaystyle}
\end{equation}
 and this condition selects the discrete complex value
\ba
\tilde E_{nlj} = \left (E_{res}\right )_{nlj} - \frac{i}{2}\Gamma_{nlj}.
\ea
 Substituting this expressions for the energy (11) and wave function (8) into
 the radial part of the Schr\"odinger equation (3), and separating the
real and
 imaginary parts, we obtain a set of two inhomogeneuos differential
equations:

\begin{equation}
\begin{displaystyle}
\left\{
\begin{array}{ccc}
\left [ -\frac{\hbar^2}{2\mu}\frac{d^2}{dr^2}+\frac{\hbar^2}{2\mu }
\frac{l(l+1)}{r^2}+V_{S.O.}+V_{Coul.} + V(r) - (E_{res})_{nlj}\right ]
u_{nlj}(r) = \frac{1}{2}\Gamma_{nlj}\cdot v_{nlj}(r), \quad & \\
\left [ -\frac{\hbar^2}{2\mu}\frac{d^2}{dr^2}+\frac{\hbar^2}{2\mu }
\frac{l(l+1)}{r^2}+V_{S.O.}+V_{Coul.} + V(r) - (E_{res})_{nlj}\right ]
v_{nlj}(r) = -\frac{1}{2}\Gamma_{nlj}\cdot u_{nlj}(r). & \\
\end{array} \right .
\end{displaystyle}
\end{equation}
 For $\Gamma_{nlj}=0$, we have $v_{nlj}(r) = 0$ and
 $\chi_{nlj}(r) = u_{nlj}(r) \equiv u^{(\circ)}_{nlj}(r)$.
 Then, the second equation vanishes and the first one converts into
homogeneous
 type as

\ba
\left [ -\frac{\hbar^2}{2\mu}\frac{d^2}{dr^2}+\frac{\hbar^2}{2\mu }
\frac{l(l+1)}{r^2}+V_{S.O.}+V_{Coul.} + V(r) \right ] u^{(\circ)}_{nlj}(r)
= E_{nlj}\cdot u^{(\circ)}_{nlj}(r).
\ea

 This homogeneous equation with the Wood-Saxon potential can not be solved by
 the ordinary method. Therefore, it should be solved via the expansion on
 the Sturm - Liouville functions [25] (the detailed description of this
method
 is given in Appendix).

 The solution of inhomogeneous differential equation of the second \\
 order can be expressed through that of the appropriate homogeneous \\
 equation $u^{(\circ)}_{nlj}(r)$ [25]:
\setcounter{tmpeqn}{0}
\arabiceqn
\ba
 u_{nlj}(r) &=& C_{u1}\cdot u^{(\circ)}_{nlj}(r)+C_{u2}\cdot
u^{(\circ)}_{nlj}(r)\int^r\frac{1}{[u^{(\circ)}_{nlj}(\eta)]^2}d\eta
 \nonumber\\
&&+u^{(\circ)}_{nlj}(r)\int^r\frac{1}{[u^{(\circ)}_{nlj}(\eta)]^2}\left
(\int^\eta\frac{2\mu}{\hbar^2}\frac{1}{2}\Gamma_{nlj}\cdot
v_{nlj}(\xi)u^{(\circ)}_{nlj}(\xi)d\xi \right )d\eta;
\ea
\ba
 v_{nlj}(r) &=& C_{v1}\cdot u^{(\circ)}_{nlj}(r)+C_{v2}\cdot
u^{(\circ)}_{nlj}(r)\int^r\frac{1}{[u^{(\circ)}_{nlj}(\eta)]^2}d\eta-
\nonumber\\
&&-u^{(\circ)}_{nlj}(r)\int^r\frac{1}{[u^{(\circ)}_{nlj}(\eta)]^2}\left
(\int^\eta\frac{2\mu}{\hbar^2}\frac{1}{2}\Gamma_{nlj}\cdot
u_{nlj}(\xi)u^{(\circ)}_{nlj}(\xi)d\xi \right )d\eta.
\ea
\reeqn
 Here all integrals are indefinte ones.
 If $\Gamma_{nlj}=0$, it follows that $u_{nlj}(r) = u^{(\circ)}_{nlj}(r)$
 and  $v_{nlj}(r) = 0$. Then, from the first equation (14.1) one obtains,
 $C_{u1} = 1$ and $C_{u2} = 0$. Similarly, $C_{v1} = C_{v2} = 0$ follows
 from the second (14.2). Therefore, we have

\setcounter{tmpeqn}{0}
\arabiceqn
\begin{eqnarray}
\qquad u_{nlj}(r) &=&
u^{(\circ)}_{nlj}(r)\nonumber\\
&&+Gu^{(\circ)}_{nlj}(r)\int^r\frac{1}{[u^{(\circ)}_{nlj}(
\eta)]^2}\left (\int^\eta v_{nlj}(\xi)u^{(\circ)}_{nlj}(\xi)d\xi \right
)d\eta; \\
v_{nlj}(r) &=&
-Gu^{(\circ)}_{nlj}(r)\int^r\frac{1}{[u^{(\circ)}_{nlj}(\eta)]^2}\left
(\int^\eta u_{nlj}(\xi)u^{(\circ)}_{nlj}(\xi)d\xi \right )d\eta, 
\end{eqnarray}
\reeqn
where  { \Large $G = \frac{2\mu}{\hbar^2}\frac{1}{2}\Gamma_{nlj}$.\\ }

Substitution of the second equation (15.2) into the first one (15.1) leads to
the integral equation for $u_{nlj}(r)$. The solution of this
integral equation is used to determine $v_{nlj}(r)$ by the second
equation (15.2). After all, one find
\setcounter{tmpeqn}{0}
\arabiceqn
\ba
u_{nlj}(r) &=&
u^{(\circ)}_{nlj}(r)-G^2u^{(\circ)}_{nlj}(r)\int^r\frac{1}{[u^{(\circ)}_{nlj
}(\eta)]^2}\int^\eta[u^{(\circ)}_{nlj}(\xi)]^2 \nonumber \\
&& \times\int^\xi\frac{1}{[u^{(\circ)}_{nlj}(\nu)]^2}\int^\nu
u^{(\circ)}_{nlj}(\tau)u_{nlj}(\tau)d\tau d\nu d\xi d\eta;
\\
v_{nlj}(r) &=&
-Gu^{(\circ)}_{nlj}(r)\int^r\frac{1}{[u^{(\circ)}_{nlj}(\eta)]^2}\left
(\int^\eta u^{(\circ)}_{nlj}(\xi)u_{nlj}(\xi)d\xi \right )d\eta.
\ea
\reeqn
 These integral equations can be solved by means of an iteration procedure\\
 (of consistent approximations): the initial approximation is chosen as
\begin{equation}
\begin{displaystyle}
\left\{
\begin{array}{l}
u^{(\circ)}_{nlj}(r) - {\rm the\ solution\ of\ the\ Sturm - Liouville\
problem;} \\
v^{(\circ)}_{nlj}(r) =
0;
\end{array} \right .
\end{displaystyle}
\end{equation}
and, then, the first iteration gives
\begin{equation}
\begin{displaystyle}
\left\{
\begin{array}{l}
u^{(1)}_{nlj}(r) = u^{(\circ)}_{nlj}(r) + \Delta\\
u^{(\circ)}_{nlj}(r);v^{(1)}_{nlj}(r) = \Delta
v^{(\circ)}_{nlj}(r);
\end{array} \right .
\end{displaystyle}
\end{equation}
where
\setcounter{tmpeqn}{0}
\alpheqn
\ba
\Delta u^{(\circ)}_{nlj}(r) &=&
-G^2u^{(\circ)}_{nlj}(r)\int^r\frac{1}{[u^{(\circ)}_{nlj}(\eta)]^2}\int^\eta
[u^{(\circ)}_{nlj}(\xi)]^2\int^\xi\frac{1}{[u^{(\circ)}_{nlj}(\nu)]^2}\nonumber\\
&& \times \int^\nu [u^{(\circ)}_{nlj}(\tau)]^2d\tau d\nu d\xi d\eta; \\
\Delta v^{(\circ)}_{nlj}(r) &=&
-Gu^{(\circ)}_{nlj}(r)\int^r\frac{1}{[u^{(\circ)}_{nlj}(\eta)]^2}\int^\eta
u^{(\circ)}_{nlj}(\xi)u^{(1)}_{nlj}(\xi)d\xi
d\eta.
\ea
\reeqn
The second iteration yields
\begin{equation}
\begin{displaystyle}
\left\{
\begin{array}{l}
u^{(2)}_{nlj}(r) = u^{(1)}_{nlj}(r) + \Delta
u^{(1)}_{nlj}(r); \\
v^{(2)}_{nlj}(r) = v^{(1)}_{nlj}(r) + \Delta
v^{(1)}_{nlj}(r);
\end{array} \right .
\end{displaystyle}
\end{equation}
where
\setcounter{tmpeqn}{0}
\alpheqn
\ba
\Delta u^{(1)}_{nlj}(r) &=&
G^2u^{(\circ)}_{nlj}(r)\int^r\frac{1}{[u^{(\circ)}_{nlj}(\eta)]^2}\int^\eta[u^{(\circ)}_{nlj
}(\xi)]^2\int^\xi\frac{1}{[u^{(\circ)}_{nlj}(\nu)]^2}\nonumber \\
&&\times\int^\nu
u^{(\circ)}_{nlj}(\tau)\cdot\Delta u^{(\circ)}_{nlj}(\tau)d\tau d\nu d\xi
d\eta; \\
\Delta v^{(1)}_{nlj}(r) &=&
-Gu^{(\circ)}_{nlj}(r)\int^r\frac{1}{[u^{(\circ)}_{nlj}(\eta)]^2}\int^\eta
u^{(\circ)}_{nlj}(\xi)\cdot\Delta u^{(1)}_{nlj}(\xi)d\xi
d\eta;
\ea
\reeqn
and etc.

 The boundary conditions (9) and (10) can be rewritten as
\setcounter{equation}{8}
\setcounter{tmpeqn}{0}
\alpheqn
\begin{eqnarray}
u_{nlj}(r)& \stackrel{r\to\infty}{\longrightarrow}&\frac{1}{2}\exp{(\tilde
k_Ir+\eta\frac{\psi}{2})}\sin{(\tilde k_Rr - \eta
\ln\frac{2\sqrt{2\mu\rho}}{\hbar}r - \frac{l\pi }{2}+
\delta_{l})}, \\
v_{nlj}(r)& \stackrel{r\to\infty}{\longrightarrow}& -\frac{1}{2}\exp{(\tilde
k_Ir+\eta\frac{\psi}{2})}\cos{(\tilde k_Rr - \eta
\ln\frac{2\sqrt{2\mu\rho}}{\hbar}r - \frac{l\pi }{2}+ \delta_{l})}, 
\end{eqnarray}
\reeqn
\setcounter{tmpeqn}{0}
\alpheqn
\begin{eqnarray}
\lim_{r\to\infty}\{[u^\prime_{nlj}(r)u_{nlj}(r) &+&
v^\prime_{nlj}(r)v_{nlj}(r)]/[u^2_{nlj}(r) + v^2_{nlj}(r)]\} = \tilde
k_I,\\
\lim_{r\to\infty}\{[u_{nlj}(r)v^\prime_{nlj}(r) &-&
u^\prime_{nlj}(r)v_{nlj}(r)]/[u^2_{nlj}(r) + v^2_{nlj}(r)]\} = \tilde
k_R.
\end{eqnarray} 
\reeqn
\setcounter{equation}{21}

 In these expressions, $\tilde k = \tilde k_R - i\tilde k_I$ and
\\
\ba
\tilde k_R =\frac{\sqrt{2\mu\rho}}{\hbar}\cos{\frac{\psi}{2}}, \qquad
\tilde k_I =\frac{\sqrt{2\mu\rho}}{\hbar}\sin{\frac{\psi}{2}},
\ea
\ba
\rho = \sqrt{(E_{res})^2_{nlj} + \Gamma^2_{nlj}/4}, \qquad tg\psi =
\frac{\Gamma_{nlj}}{2(E_{res})_{nlj}}.
\ea

 The inverse transformation gives
\\
\ba
(E_{res})_{nlj} = \frac{\hbar^2}{2\mu}(\tilde k^2_R - \tilde k^2_I),\qquad
\Gamma_{nlj} = \frac{\hbar^2}{\mu}\tilde k_R\cdot \tilde k_I.
\ea

 Thus, by having $u^{(\circ)}_{nlj}(r)$ and $(E_{res})_{nlj}$ from the
solution
 of the Sturm - Liouville problem, the wave functions $u_{nlj}(r),
v_{nlj}(r)$
 and the width $\Gamma_{nlj}$ can be calculated by means of the iteration
 procedure. And, this iteration process must converge under the execution of
 the boundary conditions (9a,b) and (10a,b).

\newpage

\begin{center}
\vspace*{0.5cm}
{ \large\bf III. RESULTS AND DISCUSSION }
\vspace*{0.3cm}
\end{center}

\ \ \ The above formalism has been used for calculation of the wave
functions $u_{nlj}(r),\ v_{nlj}(r)$, the resonance energy $(E_{res})_{nlj}$
and the width $\Gamma_{nlj}$ for $2p_{j=1/2}$ - state of neutron in $^{13}C$
nucleus. For this case, the Coulomb interaction does not play any role, and, 
thereby, it should disappear from the formalism given in the previous section. 
 
 For the Woods - Saxon potential the following parameters were used:\\
 $r_{0} = 1.46\ fm,\ a = 0.67\ fm,\ V_{0} = 32.3\ MeV\
 (E_{bind}(1p_{1/2}) = -4.947\ MeV)$.
 The wave function $u^{\circ}_{nlj}(r)$ and the energy $(E_{res})_{nlj}$ were
 obtained by means of solving the Sturm - Liouville problem [16,17]
 (see Appendix). The number of the expansion $M$ in sum (A.12) was taken
equal
 to $30$ so as to reach the enough good convergence on energy
$|E^{(M)}_{2p_{1/2}} - E^{(M-1)}_{2p_{1/2}}|\leq 0.1\ MeV$.
 The values $E^{(M)}_{2p_{1/2}}$ for the different values $M$
 are shown in Fig. 1. The value $E_{2p_{1/2}}\ =\ 8.1\ MeV$ was
 obtained at $M=30$. The value of the expanding coefficients
 $d_{n1_{1/2}}$ for the neutron wave function
 $u^{(\circ)}_{2p_{1/2}}(r)$ are presented in Table 1.

 For $\Gamma_{n_{c}lj} = \Gamma_{2p_{1/2}} = 22.244\ MeV$, the wave functions
 $u_{2p_{1/2}}(r)$ and $v_{2p_{1/2}}(r)$ after 5th iteration were tailed
 smoothly together with their asymptotical expressions, i.e. the boundary
 conditions (9a,b) and (10a,b) were fulfilled.

 For the time dependent part of wave function, we have
\ba
\qquad\qquad\nonumber \\
\Psi(t) = exp{\left (-\frac{i}{\hbar}\tilde E_{nlj}\cdot t\right )} =
exp{\left (-\frac{i}{\hbar}(E)_{nlj}\cdot t\right )}\cdot exp{\left
(-\frac{\Gamma_{nlj}}{2\hbar}\cdot t\right )} = \nonumber\\
\qquad\qquad\nonumber \\
= exp{\left (-i\frac{\hbar}{2\mu}(\tilde k^2_R - \tilde k^2_I)\cdot
t\right )}\cdot exp{\left (-\frac{\hbar}{2\mu}\tilde k_R\cdot \tilde
k_I\cdot t\right )},
\ea
 where the first factor is the wave function oscillating over time,\\
 and the second factor is the descending function over time.\\
 There were used relations (11) and (24).

Notice that the full radial wave function is
\ba
\tilde \Psi_{nlj}(r,t) = \chi_{nlj}(r)\cdot\Psi(t).
\ea

 If this wave function is considered in the 3 - dimensional space
$(z,r,t)$,\\
 where $\ z={ Re} \tilde \Psi_{nlj}(r,t)$ or
 ${ Im} \tilde \Psi_{nlj}(r,t)$, then one obtains a complicated surface
 of the second order.\\
 In the plane $(z,r)$, at $t=0$, the $u_{nlj}(r)$ and $v_{nlj}(r)$ are
 oscillating and increasing in amplitude (i.e. diverging)
 when  $r \to \infty$. In the plane $(z,t)$, at $r=0$, the real and
 imaginary parts of $\chi_{nlj}(r=0)\cdot \Psi(t)$ are oscillating and
 decreasing in amplitude (i.e. convergent) when $t\to \infty$.
 And if the full wave function is considered in the plane $(z,t(r))$ for
 $t=r/v$, where $v=\frac{\hbar k_N}{\mu}$ is nucleon speed at the resonance
 energy $(E_{res})_{nlj}$ of $2p_{1/2}$ - state
 ($k_N = \frac{\sqrt{2\mu(E_{res.})_{nlj}}}{\hbar}$ --
 the wave vector for nucleon in  $2p_{1/2}$ - state), then
\ba
\qquad\qquad\nonumber \\
\tilde \Psi(r,t(r)) = \chi_{nlj}(r)\cdot exp{\left (-i\frac{(\tilde k^2_R
- \tilde k^2_I)}{2k_N}\cdot r\right )}\cdot exp{\left (-\frac{\tilde
k_R\cdot \tilde k_I}{k_N}\cdot r \right )}.
\ea
\\
 Such a wave function at $r \to \infty$ is convergent and can be
 normalized to {\bf1}. The Fig. 2 shows the real and imaginary parts
 of the normalized radial wave function for the neutron in
 $\left (2p_{1/2}\right )$ - state of $^{13}C$.
 With such wave functions, the root-mean-square radius (r.m.s.),
 $<r^{2}_{2p_{1/2}}>^{\frac{1}{2}}\ =\ 11.92\ fm$,
 for $2p_{1/2}$ - state of neutron was calculated.
 For the neutron in $1p_{1/2}$ - state, we have
 $<r^{2}_{1p_{1/2}}>^{\frac{1}{2}}\ =\ 3.40\ fm$. The value of
 the r.m.s. radius of neutron in $2p_{1/2}$ - state is much larger than
 one in $1p_{1/2}$ -state. Therefore, the neutron $2p_{1/2}$ - state may be
 regarded as the "halo - like" state.

 Generally, the resonance wave function obtained from the time-independent
Schr\"{o}dinger equation is exponentially divergent at an asymptotic distance.
Therefore, it is not easy to calculate the r.m.s. radius with such an unnormalizable 
resonance state wave function. One way to calculate the r.m.s. radius is to use the
complex scaling method[22] or to introduce a convergence factor such as 
$\exp^{-\alpha r^{2}}$ provided $\alpha\to\infty$ at the end[23].
It should be noticed that the r.m.s. radius calculated by the last method[23],
i.e. convergence factor method, is a complex number for the resonance state, e.g.
$<r^{2}_{3S1/2}>=9.41+i4.52fm$ for the  overbarrier resonance state $3S_{1/2}$ with
$E_{r} =1.63MeV$ and $\Gamma =0.246MeV$ (the barrier top is 1.50 MeV).
Consequently, the quadrupole moments of nuclei will be complex values, too,
in their approach. This sequence seems to require further consideration and deeper 
investigation.
 On the other hand, our method of calculation for the r.m.s. radius takes into account 
the time-dependent part of wave function by means of $t=\mu r/\hbar k_{N}$. As is 
explained  above, this time-dependent part makes the full wave function convergent.
Thus, the r.m.s. radius can be calculated without any difficulty.  

\newpage

\begin{center}
\vspace*{0.5cm}
{ \large\bf IV. CONCLUSION }
\vspace*{0.3cm}
\end{center}

 In this paper, we suggested the method of solving the system of
 inhomogeneous differential equations of the second order with help of
 preliminary determined solution of corresponding homogeneous equation.
 Such an approach allows to obtain the wave functions, energy and width for
 the overbarrier resonance state.

 This method was used for analyzing $2p_{1/2}$ - resonance
 state of the neutron in $^{13}C$ nucleus.
 The value of root-mean-square radius of neutron in this state is larger
 than that in the $1p_{1/2}$ - state.

 In the framework of this method, we propose to calculate neutron and proton
 resonance $2p$ - states in  $^{13}C$ and $^{13}N$ nuclei and then to explore
 them for analyzing the electron scattering data and
 $\beta$ - transition. It can also be a good application of our method to
 analyze the elastic proton scattering from $^{13}C$ at 1 GeV [27].

 Besides, we suggest in a new fashion to look at the structure of
 the halo - nuclei, in which an existence of the sub - and over - barrier
 resonant single-particle states, as well as the correlated multi-particle
 states can be displayed.

\begin{center}
\vspace*{0.5cm}
{ \large\bf ACKNOWLEDGMENTS }
\vspace*{0.3cm}
\end{center}
     We are grateful to R.A. Eramzhyan and F.A. Gareev  for
        interest to this work and very helpful scientific discussions.
  This work was supported by KOSEF(961-0204-018-2) and the Korean Ministry of Education
        (BSRI 97 - 2425). G. Kim thanks KOSEF 
  for financial support while visiting Yonsei University.

%\newpage

\begin{center}
\vspace*{0.5cm}
{ \large\bf APPENDIX A }
\vspace*{0.3cm}
\end{center}

 The common theory of the expansion on the Sturm - Liouville functions was
 developed long ago [28,29,30]. It is actually the generalization of
 the expansion in the Fourier series. Examples of application of this
theory to
 the atomic and nuclear problems are given in [25].
 The  Sturm - Liouville equation for eigenvalues $\alpha_{nlj}$ and
 eigenfunction $\varphi_{nlj}(r)$ is

\ba
\left [ -\frac{\hbar^2}{2\mu}\left (
\frac{d^2}{dr^2}-\frac{l(l+1)}{r^2}\right ) + V_{S.O.} + V_{Coul.} +
\alpha_{nlj}\cdot V(r) \right ] \varphi_{nlj}(r) =
E_{0lj}\varphi_{nlj}(r),(A1) \nonumber
\ea
\\
 where $E_{0lj}$ is the fixed negative number (in our case this is the
binding
 energy of nucleon in $1p$ - shell).\\
 The functions $\varphi_{nlj}(r)$  must satisfy the following boundary
 condition,

\ba
\qquad\qquad\qquad\qquad \varphi_{nlj}(r) \to 0 \quad\mbox{at}\quad r\to 0
\quad \mbox{and}\quad r \to \infty. \quad\qquad\qquad\ \ \ \ \ \ \nonumber
\ea
\\
 If $\alpha_{0lj}=1$ and $E_{0lj}$ is the eigenvalue of the Schr\"odinger
 equation, then the solution of this equation will be the first
 eigenfunction of the Sturm - Liouville problem, i.e. $ \varphi_{0lj}(r) $ is
 the ordinary physical wave function.
 For fixed values $l$ and $j$, the depth of potential will grow with
increasing
 $\ \alpha_{nlj}\ $. If we find such $\alpha_{nlj}$, that energy of
 $n$ - state coincides with $E_{0lj}$, then the obtained wave function
 will be the eigenfunction of the Sturm - Liouville problem.
 The eigenvalues $\alpha_{nlj}$ are infinite succession of positive
 discrete numbers, which satisfy the condition:

\ba
\alpha_{0lj}<\alpha_{1lj}<\alpha_{2lj}< ... \alpha_{n-1,lj}< \alpha_{n,l}
< ... \nonumber
\ea
\\
 and $\alpha_{nlj}\to \infty $ at $n\to \infty $.
 It follows from the Hermite conjugate of the Hamiltonian operator for
 the Sturm - Liouville problem that the functions $\varphi_{nlj}(r)$  are
 orthogonal with the weight $V(r)$ at the fixed $l,j$ and $E_{0lj}$, i.e.

 \ba
\qquad\qquad\qquad\qquad\qquad
\int^{\infty}_{0}\varphi_{n^{'}lj}V(r)\varphi_{nlj}(r)dr =
-\delta_{n^{'}n}. \qquad\qquad\ \ \ \ \ \ \ \ \ \ \ \ \ \ (A2) \nonumber
 \ea
\\
 Other mathematical aspects of this method (e.g. the problems of sign of\\
 the eigenvalues, of the completeness and the convergence of the expansion at
 $r\to \infty$) are considered in works [25] in detail.

 It is very convenient to present the eigenfunctions $\varphi_{nlj}(r)$
 in an analytical form with parameters calculated by means of modifying
 quasi-classical method [31]:

\ba
\quad\qquad\qquad\qquad\qquad \varphi_{nlj}(r) =
N_{nlj}(S^{'})^{-\frac{1}{2}}exp{\left (-\frac{S^2}{2}\right ) }H_n(S),\ \
\ \ \ \ \ \ \ \ \ \ \ \ \ \ \ \ \ (A3) \nonumber
\ea

 where $N_{nlj}$ is coefficient of norm, $H_{n}(S)$ is Hermite polynomial,
 $S'\ =\ dS/dr$, and $S(r)$ is the correct function to be found by means of
 numerical calculations.

 As the comparison equation, the equation of the harmonic oscillator type
 is chosen,

\ba
\quad\qquad\qquad\qquad\qquad\qquad \frac{d^2\Phi}{dS^{2}}+
(2n+1-S^{2})\Phi = 0, \qquad\ \ \ \ \ \ \ \ \ \ \ \ \ \ \ \ \ \ \ \ \ \
(A4) \nonumber
\ea

and the integral relation is obtained for the definition $S(r)$ [25]:

\ba
\quad\qquad\qquad\qquad\qquad
\int^{S}_{-\sqrt{2n+1}}(2n+1-\sigma^2)^{\frac{1}{2}}d\sigma =
\int^{r}_{r_{1}} p(\xi ) d\xi , \qquad\qquad\ \ \ \ \ \ \ \ (A5) \nonumber
\ea

 where

\ba
\qquad\qquad p(r) = \sqrt{\frac{2\mu}{\hbar^2}(E_{nlj}-\alpha_{nlj}V(r))-
V_{S.O.} - V_{Coul.} - \frac{(l+\frac{1}{2})^2}{r^{2}}} \quad\ \ \ \ \ \ \
\ \ (A6) \nonumber
\ea

 is the momentum analogous to the quasi-classical one, $r_{1}$ is the least
 root of equation $p(r)=0$.

 In our work, we express $S(r)$ in term of

{ \large
\ba
\quad\qquad\qquad\qquad
S(r)=\frac{A_o}{2}+\sum^{k_{max}}_{k=1}A_kT_k\biggl (
\frac{2r-a-b}{b-a}\biggr ),\qquad\qquad\quad \ \ \ \ \ (A7) \nonumber \\
\quad\qquad\qquad\qquad
A_k=\frac{2}{k_{max}+1}\sum^{k_{max}}_{j=0}S(r_j)\cos{\biggl (
\frac{(2j+1)k\pi}{2k_{max}+2}\biggr )},\qquad\qquad\quad \ \ \ \ \ (A8)
\nonumber \\
\quad\qquad\qquad\qquad r_j=\frac{a+b}{2}+ \frac{b-a}{2}\cos{\biggl (
\frac{(2j+1)\pi}{2k_{max}+2}\biggr )},\qquad\qquad\quad \ \ \ \ \ \ (A9)
\nonumber
\ea
 }

 where $T_k(\xi )$ is Chebyshev polynomial of $k$ power;
 $a=r_{1}(E_{nlj})$ and $b=r_{2}(E_{nlj})$ are the turning points, which are
 defined by the equation:

\ba
\quad\qquad\qquad E_{nlj}-\alpha_{nlj}\cdot V(r) -
V_{S.O.}-V_{Coul.}-\frac{\hbar^{2}}{2\mu
}\frac{(l+\frac{1}{2})^{2}}{r^2}=0,\qquad \ \ \ \ \ \ (A10) \nonumber
\ea
\\
$k_{max}$ is maximum number of expansion (in our case, $k_{max}=30$ was took
for reaching the necessary uniformity).

 Having such a Sturm - Liouville functions, we can solve the homogeneous
 Schr\"odinger equation

\ba
\left [ -\frac{\hbar^2}{2\mu}\frac{d^2}{dr^2}+\frac{\hbar^2}{2\mu }
\frac{l(l+1)}{r^2}+V_{S.O.}+V_{Coul.} + V(r) \right ] u^{(\circ)}_{nlj}(r)
= E_{nlj}\cdot u^{(\circ)}_{nlj}(r)\ \ \ \ \ (A11) \nonumber
\ea
\\
 with the boundary condition $\chi_{nlj}\to 0 $ for $r\to 0$ and $r\to
\infty $,
 by means of expansion $u^{(\circ)}_{nlj}(r)$ with respect to
 $\varphi_{nlj}(r)$ :

 \ba
\quad\qquad\qquad\qquad\qquad\qquad u^{(\circ)}_{nlj}(r)=\sum^{\infty
}_{n^{'}=0}d_{n^{'}lj}\cdot \varphi_{n^{'}lj}(r).\qquad\qquad\qquad\quad \
\ \ \ \ \ (A12) \nonumber
 \ea

 If we substitute this expansion into equation (A11), add and subtract
 $\alpha_{nlj}V(r)$, multiply on left-side to $\varphi_{nlj}(r)V(r)$, and
 carry out integration over $r$, then the following equation can be obtained,

 \ba
\quad\qquad (E - E_{0lj})d_{nlj}+\sum^{\infty
}_{n^{'}=0}d_{n^{'}lj}(1-\alpha_{n^{'}lj }) \int\varphi_{nlj}V^{2}(r)
\varphi_{n^{'}lj}dr = 0,\quad \ \ \ \ \ \ \ (A13) \nonumber
 \ea
 \\
 which is an infinite set of equations. In order to solve this set of
 equations, it is necessary to cut off the summation at some fixed
 number $M$. Then, from the condition for solving this set of equations, one
 finds the approximate eigenvalue $E^{(M)}_{nlj} \equiv(E_{res})_{nlj}$ and
 the coefficients $d^{M}_{nlj}$, accordingly the eigenfunction
 $u^{(\circ)(M)}_{nlj}(r)$.\\
 \\
 At $M\to \infty,\quad E^{(M)}_{nlj}\to E^{exact}_{nlj}
\equiv(E_{res})_{nlj}$.

\newpage

\bb{99}
\bi{1} E.K. Warburton, D.E. Alburger and D.J Millener,\\
\hspace*{0.1cm}Phys. Rev. C{\bf22}, 2330 (1980).\\
\vspace*{-0.7cm}
\bi{2}T. Suzuki, H. Hyuga, A. Arima and K. Yazaki,\\
\hspace*{0.1cm}Nucl. Phys. A{\bf358}, 421 (1981); Phys. Lett. B{\bf106},
19 (1981).\\
\vspace*{-0.7cm}
\bi{3}R.S. Hicks, J. Dubach, R.A. Lindgren et al.,\\
\hspace*{0.1cm}Phys. Rev. C{\bf26}, 339 (1982).\\
\vspace*{-0.7cm}
\bi{4}P.W.M. Glaudemans,\\
\hspace*{0.1cm}{\it Symmetries in Nuclear Structure} (Plenum, New York,
1983);\\
\hspace*{0.1cm}{\it Nuclear Shell Models} (World-Scientific, Singapore,
1985).\\
\vspace*{-0.7cm}
\bi{5}T.W. Donnelly, I. Sick, Rev. Mod. Phys. {\bf56}, N3, 461 (1984).\\
\vspace*{-0.7cm}
\bi{6} P.G. Blunden and B. Castel, Nucl.Phys. A{\bf445}, 742 (1985).\\
\vspace*{-0.7cm}
\bi{7}R.S. Hicks, R.L. Huffman, R.A. Lindgren et al.,\\
\hspace*{0.1cm}Phys. Rev. C{\bf36}, 485 (1987).\\
\vspace*{-0.7cm}
\bi{8}A.G.M. van Hees, A.A. Wolters and P.W.M. Glaudemans,\\
\hspace*{0.1cm}Nucl. Phys. A{\bf476}, 61 (1988).\\
\vspace*{-0.7cm}
\bi{9}D.J. Millener, D.I. Sober, H. Crannell et al.,\\
\hspace*{0.1cm}Phys. Rev. C{\bf39}, 14 (1989).\\
\vspace*{-0.7cm}
\bi{10}K. Amos, L. Berge and D. Kurath, Phys. Rev. C{\bf40}, 1491 (1989).\\
\vspace*{-0.7cm}
\bi{11}A.A. Wolters, A.G.M. van Hees and P.W.M. Glaudemans,\\
\hspace*{0.1cm}Phys. Rev. C{\bf42}, 2053; C{\bf42}, 2062 (1990); C{\bf45}, 477 (1992).\\
\vspace*{-0.7cm}
\bi{12}C. Benhold and L. Tiator, Nucl. Phys. A{\bf159}, 805 (1990);\\
\hspace*{0.1cm}Phys. Lett. B{\bf328}, 31 (1990).\\
\vspace*{-0.7cm}
\bi{13}S.S. Kamalov, C. Benhold, P. Mach, Phys. Lett. B{\bf259}, 410
(1990).\\
\vspace*{-0.7cm}
\bi{14}Ye. Ismatov, G. Kim, A.V. Khugaev et al.,\\
\hspace*{0.1cm}Ukrain. Jour. of Phys. {\bf40}, N3-4, 414 (1995).\\
\hspace*{0.1cm}G. Kim, A.V. Khugaev, Il-T. Cheon, M.T. Jeong,\\
\hspace*{0.1cm}(submitted to Z.Phys.A.)\\
\vspace*{-0.7cm}
\bi{15}A.A. Chumbalov, S.S. Kamalov, R.A. Eramzhan,\\
\hspace*{0.1cm}Nucl. Phys. A{\bf581}, 543 (1995).\\
\newpage
\vspace*{-0.7cm}
\vspace*{0.3cm}
\bi{16}F.A. Gareev, G. Kim, A.V. Khugaev,\\
\hspace*{0.1cm}Preprint JINR, ø4-95-434, Dubna (1995).\\
\hspace*{0.1cm}Yad. Fiz. {\bf12}, 1240 (1997).\\
\vspace*{-0.7cm}
%\newpage
%\vspace*{0.3cm}
\bi{17}G. Kim, A.V. Khugaev,\\
\hspace*{0.1cm}Izv. Acad. Sci. of Russia, ser.phys., {\bf60}, N5, 123;
{\bf60}, N11, 107 (1996).\\
\vspace*{-0.7cm}
\bi{18}V. Gillet and N. Vinh Mau,\\
\hspace*{0.1cm}Nucl. Phys. {\bf54}, 321 (1964).\\
\vspace*{-0.7cm}
\bi{19}V.D. Mur, V.S. Popov,\\
\hspace*{0.1cm}JETP(Journal of Exper. and Theor. Phys.) {\bf104}, 2293
(1993);\\
\hspace*{0.1cm}Phys. Lett. A{\bf157}, 185 (1991).\\
\hspace*{0.1cm}B.M. Karnakov, V.D. Mur, S.G. Pozdnyakov, V.S. Popov,\\
\hspace*{0.1cm}Yad. Fiz. {\bf54}, 400 (1991).\\
\vspace*{-0.7cm}
\bi{20}A.M. Lane, R.G. Thomas,\\
\hspace*{0.1cm}Rev. Mod. Phys. {\bf30}, 257 (1958).\\
\vspace*{-0.7cm}
\bi{21}G. Garcia-Calderon,\\
\hspace*{0.1cm}Nucl. Phys. A{\bf261}, 130 (1976);\\
\hspace*{0.1cm}G. Garcia-Calderon, R. Peierls,\\
\hspace*{0.1cm}Nucl. Phys. A{\bf265}, 443 (1976).\\
\vspace*{-0.7cm}
\bi{22}B.Gyarmati and T.Vertse,\\
\hspace*{0.1cm}Nucl.Phys. A{\bf160}, 523 (1971).\\
\vspace*{-0.7cm}
\bi{23}M. Homma, T. Myo and K. Kat\=o,\\
\hspace*{0.1cm}Prog. Theor. Phys. {\bf97}, 561 (1997).\\
\vspace*{-0.7cm}
\bi{24}A.I. Bazj, Ya.B. Zel'dovich, A.M. Perelomov,\\
\hspace*{0.1cm}{\it Scattering, reactions and decays in nonrelativistic
quantum mechanics}\\
(Moscow, Nauka, 1971).\\
\vspace*{-0.7cm}
\bi{25}F.A. Gareev, S.P. Ivanova, N.Ju. Shirikova,\\
\hspace*{0.1cm}Sov. Jour. Theor. and Math. Phys. {\bf8}, N1, 97 (1971).\\
\hspace*{0.1cm}J.M. Bang, F.A. Gareev, S.P. Ivanova,\\
\hspace*{0.1cm}Fiz. Elem. Chastits At. Yadra {\bf9}, 286 (1978) [Sov. J.
Part. Nucl.].\\
\vspace*{-0.7cm}
\bi{26}E. Kamke, {\it Diffrentialgleichungen losungsmethoden und losungen}\\
\hspace*{0.1cm}(Leipzig, 1959).\\
\vspace*{-0.7cm}
\bi{27}G.D. Alkhazov, {\it Izv. Akad. Nauk SSSR, Ser. Fiz.}\\
\hspace*{0.1cm}[Bull. Acad. Sci. SSSR. Phys.Ser.], {\bf42}, No.11, 2218
(1978).\\
\vspace*{-0.7cm}
\bi{28}J.C.F. Sturm, {\it Memories presents des Sovants Etrangers.}\\
\hspace*{0.1cm}1935, V. 6, P. 271;\\
\hspace*{0.1cm}J. Liouville, J. Math., 1846, V. 11, P. 221.\\
\vspace*{-0.7cm}
\bi{29}R. Courant, D. Hilbert, {\it Methods of Mathematical Physics.}\\
\hspace*{0.1cm}N.Y. Internat. Science Publ., 1953.\\
\vspace*{-0.7cm}
\bi{30}E.C. Titchmarsh, {\it Eigenfunction Expansion Associated with
Second-Order\\
\hspace*{0.1cm}Differential Equations.} Oxford University Press, 1958.\\
\vspace*{-0.7cm}
\bi{31}S.C. Miller, R.H. Good, Phys. Rev. {\bf91}, 91 (1953);\\
\hspace*{0.1cm}B.N. Kalinkin, Ja. Grabovsky, F.A. Gareev,\\
\hspace*{0.1cm}Acta Phys. Polonica XXX, 999 (1966).\\
\eb

\newpage
\vspace*{1.cm}
\begin{center}
 TABLES\\
\vspace*{0.5cm}
 TABLE 1. The values of the expanding coefficients $d_{n1_{1/2}}$ for
 the neutron wave function $u^{(\circ)}_{2p_{1/2}}(r)$ in $^{13}C$.\\
\vspace*{1.cm}

 \begin{tabular}{|c|c|c|c|c|c|} \hline \hline
 n &$d_{n1_{1/2}}$&$n$&$d_{n1_{1/2}}$&$n$&$d_{n1_{1/2}}$  \\ \hline \hline
 1 &  .67484 & 11& -.02874 & 21& -.00491\\
 2 &  .62089 & 12&  .02375 & 22&  .00408\\
 3 & -.29481 & 13& -.01974 & 23& -.00337\\
 4 &  .18248 & 14&  .01651 & 24&  .00275\\
 5 & -.12831 & 15& -.01386 & 25& -.00221\\
 6 &  .09306 & 16&  .01166 & 26&  .00174\\
 7 & -.06997 & 17& -.00984 & 27& -.00132\\
 8 &  .05489 & 18&  .00829 & 28&  .00096\\
 9 & -.04343 & 19& -.00699 & 29& -.00064\\
 10&  .03521 & 20&  .00587 & 30&  .00036\\  \hline \hline
\end{tabular}\\
\end{center}

\newpage

\vspace*{0.5cm}
\begin{center}
 FIGURES\\
\end{center}
\vspace*{0.5cm}

\indent FIG. 1. The energy values $E^{(M)}_{2p_{1/2}}$ versus the number of 
         expansion, M,for the neutron wave function, i.e. M is the number of terms 
         taken in the summation (A12).

\vspace*{0.5cm}

 FIG. 2. Radial wave function for neutron in $2p_{1/2}$ - state:\\
      a) real and  b) imaginary parts.

}
\end{document}